\documentclass[english,floatfix,twocolumn,showkeys,showpacs,superscriptaddress,nofootinbib,aps,prd]{revtex4}
\usepackage[latin1]{inputenc}
\usepackage[T1]{fontenc}
\usepackage{lmodern}
\usepackage{amsmath}
\usepackage{amssymb}
\usepackage{graphicx}
\usepackage{esint}
\usepackage{longtable}
\usepackage{dcolumn}
\usepackage{babel} 
\usepackage{csquotes} 
\usepackage{color} 

\begin{document}

\title{Black holes with a cosmological constant in bumblebee gravity}

\author{R. V. Maluf}
\email{r.v.maluf@fisica.ufc.br}
\affiliation{Universidade Federal do Ceará (UFC), Departamento de Física,\\ Campus do Pici, Fortaleza - CE, C.P. 6030, 60455-760 - Brazil}

\author{Juliano C. S. Neves}
\email{nevesjcs@if.usp.br}
\affiliation{Centro de Ciências Naturais e Humanas, Universidade Federal do ABC,\\ Avenida dos Estados 5001, Santo André, 09210-580 São Paulo, Brazil}

\begin{abstract}
In this work, we present black hole solutions with a cosmological constant in bumblebee gravity, which provides
a mechanism for the Lorentz symmetry violation by assuming a nonzero vacuum expectation value for the
bumblebee field. From the gravitational point of view, such solutions are spherically symmetric black holes
with an effective cosmological constant and are supported by an anisotropic energy-momentum tensor, conceived of as 
the manifestation of the bumblebee field in the spacetime geometry. Then we calculate the shadow angular radius
for the proposed black hole solution with a positive effective cosmological constant. In particular, our results are the 
very first relation between the  bumblebee field and the shadow angular size.  
\end{abstract}


\maketitle

\section{Introduction}

The recent search for signs of the Lorentz symmetry violation at low energy regimes, due to remaining effects 
of quantum gravity on the Planck scale, has attracted attention in the last years \cite{Liberati2013}. 
The Lorentz-violating effects arise in different contexts such as string 
theory \cite{Kostelecky1989,Kostelecky1989a,Kostelecky1989b,Kostelecky1991}, noncommutative spacetime \cite{noncommutative,noncommutative2,noncommutative3}, 
loop quantum gravity \cite{LQG,LQG2}, warped brane worlds \cite{BraneLIV,BraneLIV2}, and 
Ho$\breve{{\rm r}}$ava-Lifshitz gravity \cite{Horava2009}, among others. A suitable framework to 
account for Lorentz-violating effects on the behavior of elementary particles was proposed by Colladay and 
Kostelecký \cite{Kostelecky1997,Kostelecky1998}, based on the idea of a spontaneous Lorentz symmetry 
breaking in string theory \cite{Kostelecky1989}, known as the Standard Model extension (SME). 

The SME is an effective field theory, which includes additional gauge-invariant terms, even compatible with the 
observer Lorentz invariance, and is composed of contractions of the physical standard model fields with fixed background 
tensors \cite{Bluhm2005,Bluhm2008}. Several studies involving the different sectors of the SME were carried 
out and allowed to raise stringent bounds on the magnitude of the Lorentz-violating parameters. Theoretical and 
phenomenological developments include CPT symmetry violation \cite{CPTviolation,CPTviolation2,CPTviolation3,CPTviolation4,CPTviolation5,CPTviolation6,CPTviolation7},
 the fermion sector \cite{fermionsector,fermionsector2,fermionsector3,fermionsector4,fermionsector5,fermionsector6,fermionsector7}, 
the gauge CPT-odd/even sectors \cite{gaugesector,gaugesector2,gaugesector3,gaugesector4,gaugesector5,gaugesector6,gaugesector7,gaugesector8,gaugesector9}, photon-fermion interactions \cite{fermioninteraction,fermioninteraction2,fermioninteraction3,fermioninteraction4,fermioninteraction5,fermioninteraction6}, and radiative corrections \cite{rad1,rad1a,rad1b,rad1c,rad1d,rad1e,rad1f,rad1g,rad1h,rad1i,rad1j,rad2,rad2a,rad2b,rad2c,rad2d,rad2e}.

When the gravitational interaction is taken into account, a spontaneous symmetry breaking 
mechanism \cite{Bluhm2005,Bluhm2008} is adopted in order to implement the local Lorentz violation
and preserve the geometric constraints and conservation laws required by general relativity. 
An interesting and consistent approach to the spontaneous local Lorentz and diffeomorphism violations that includes
 the gravitational sector in the SME framework was initially discussed in Ref. \cite{Kostelecky2004}. In 
 such an approach, spacetime is assumed to be a Riemann-Cartan manifold with nonzero curvature and torsion. 
 The presence of a self-interacting potential for the tensor fields guarantees a vacuum expectation value (VEV) whose 
underlying dynamics are built from vierbein and spin connections \cite{Bailey2006}. Recent studies 
involving the SME gravitational sector include models that describe the expansion of the universe \cite{cosmology}, 
linearized gravity \cite{linearized,linearized2}, and gravitational waves \cite{GravityWaves,GravityWaves2}. 
It is worth mentioning the existence of alternative approaches that include the violation of the Lorentz symmetry directly 
in the geometric structure of the theory, such as in the Randers spacetime
 \cite{RandersSpace,RandersSpace2,RandersSpace3} and the bipartite-Finsler
 spacetime \cite{Bipartite,Bipartite2}.

The bumblebee models are examples of proposals that involve self-interacting tensor fields with a nonzero 
VEV, fields which define a privileged direction in spacetime, that is to say, they generate an anisotropic
energy-momentum tensor. The simplest case is described by a vector
 field $B_{\mu}$ and was first considered in the context of string theories \cite{Kostelecky1989}, with 
 the spontaneous Lorentz-symmetry breaking triggered by a smooth quadratic potential. The bumblebee models have 
 been studied in different contexts, whether in the curved spacetime \cite{Kostelecky2004, Bailey2006, linearized, 
 BumblebeeCurved} or in the Minkowski spacetime \cite{BumblebeeFlat,BumblebeeFlat2,BumblebeeFlat3,BumblebeeFlat4,BumblebeeFlat5,BumblebeeFlat6}. 
 As is known in the literature, some interesting effects of the bumblebee VEV arise in connection with black 
 hole physics. In this article, we will focus on some of them. 

Initial studies on black hole solutions within the Lorentz violation scenarios were carried out by 
Bertolami and Páramos  \cite{Bertolami2005}, where the authors imposed the condition of a covariant constant 
VEV, that is, $\nabla_{\mu}b_{\nu}=0$ (where $\nabla_{\mu}$ is the covariant derivative), 
instead of the common prescription $\partial_{\mu}b_ {\nu}=0$.
By assuming different configurations for the background vector, those authors 
obtained approximated solutions in terms of the parametrized post-Newtonian parameters that are
modified by the presence of the bumblebee VEV. A static and spherically symmetric black hole 
solution was recently built by Casana \textit{et al.} \cite{Casana2018}, by considering a nonminimal coupling between 
the Ricci  tensor and the bumblebee field. The additional condition for the constant squared norm of the VEV 
allowed them to find an exact Schwarzschild-like solution. Modifications on the black hole thermodynamics 
due to the presence of the nonvanishing bumblebee VEV were pointed out in Ref. \cite{Debora2020}, 
considering black holes geometries obtained in Refs. \cite{Bertolami2005,Casana2018}. Other black hole
solutions were obtained involving different bumblebee models. A Kerr-like solution was built following a 
similar approach \cite{Casana2020}. In Ref. \cite{Euclides2020}, a Reissner-Nordström solution 
emerged from the spontaneous Lorentz symmetric breaking triggered by a Kalb-Ramond field. Even exotic 
wormhole solutions have been investigated in the literature  of that context \cite{Ovgun2020,Rondineli2019}.
 
Having said all that, we follow some mentioned works and present new black holes geometries
with a cosmological constant in the bumblebee model or gravity.  As will see, such solutions with a 
cosmological constant are just possible by assuming a suitable form for the bumblebee potential. 
However, as we will see, the proposed geometries here are neither asymptotically anti-de Sitter
 nor asymptotically de Sitter. In this sense, we call them Schwarzschild-anti-de Sitter-like and Schwarzschild-de Sitter-like
 black holes, as much as the zero cosmological constant case, obtained by Casana \textit{et al.} \cite{Casana2018}, 
 is not asymptotically flat, thus it is called Schwarzschild-like black hole. Studies on black holes with a cosmological 
 constant are considered as very important issues since the
 anti-de Sitter/conformal field theory (AdS/CFT) correspondence and the observation of the accelerating expansion
 of the universe (in the latter case, asymptotically de Sitter or de Sitter-like black holes are justified). 
 
From the metric obtained here, we calculated a very important observable in order to relate the bumblebee field to
the spacetime geometry: the shadow angular radius. Shadow of black holes has been a seminal topic in 
physics recently. And the reason for that is the very first image of a black hole 
announced by the Event Horizon Telescope Collaboration in 2019 \cite{EHT,EHT2}. 
 Indeed, that famous image shows the shadow 
of M87*, the central supermassive black hole in the Messier 87 galaxy. However, the very  first shadow of a black hole
was calculated in the last century. In the 60s Synge \cite{Synge} obtained that which we call today 
shadow of the Schwarzschild black hole. Then Bardeen did the same for the Kerr geometry \cite{Bardeen}.
 In the recent years, shadows have been drawn
for several black holes in many contexts \cite{Eiroa,Neves1,Neves2,Vagnozzi,Khodadi,Casana2020}.
 According to recent works, relations between the shadow and  the black hole
parameters are possible in the general relativity realm or even in contexts beyond the Einsteinian
context \cite{Neves1,Neves2,Vagnozzi,Kumar,Khodadi}. Our focus here
is in a model beyond general relativity. 

This article is structured as follows: In Section II the framework is presented, and both the Schwarzschild-anti-de Sitter-like and the
Schwarzschild-de Sitter-like black holes are built 
in the bumblebee gravity,  some features are discussed like horizons, singularity, and the energy-momentum
tensor for those black holes. Section III speaks of the shadow angular radius of the Schwarzschild-de Sitter-like 
geometry, and the influence of the Lorentz-violating parameter on this phenomenon is pointed out. 
The final comments are in Section IV. We adopt geometrized units in our calculations, i.e.,
$G=c=1$, where $G$ is the Newtonian constant, and $c$ is speed of light in vacuum.

\section{Constructing black holes in the bumblebee gravity}

\subsection{The adopted framework}

As we pointed out, bumblebee models provide a simple mechanism for studying the spontaneous 
breaking of the Lorentz symmetry in the gravitational scenario. These types of models have a nontrivial 
VEV  that affects the dynamics of other fields coupled to the bumblebee field, 
 preserving geometric structures and conservation laws compatible with a usual pseudo-Riemannian manifold
  from general relativity \cite{Kostelecky2004,Bluhm2005}. 

Among several possibilities of  models that are able to break the Lorentz symmetry, there is the simplest action form 
involving a vector field $B_{\mu}$, the bumblebee field, in a torsion-free spacetime  written as
\begin{eqnarray}
 S_{B}&=&\int d^{4}x\sqrt{-g} \bigg [  \frac{1}{2\kappa}\left(R-2\Lambda\right)+\frac{\xi}{2\kappa}B^{\mu}B^{\nu}R_{\mu\nu} \nonumber \\ 
 &&-\frac{1}{4} B_{\mu\nu}B^{\mu\nu}-V(B^{\mu}B_{\mu}\pm b^{2})+\mathcal{L}_{M} \bigg ], 
 \label{S1} 
\end{eqnarray}
 where $\kappa=8\pi G/c^4$ is the gravitational coupling constant, $\Lambda$ is the cosmological 
 constant, and $\xi$ plays the role of a coupling constant that accounts for the nonminimum interaction 
 between the bumblebee field and the Ricci tensor or geometry (with mass 
 dimension $[\xi]=M^{-1}$) \cite{Bailey2006,linearized}. Also, one has
  $B_{\mu\nu}\equiv\partial_{\mu}B_\nu-\partial_{\nu}B_{\mu}$ or the bumblebee field strength, 
  and $\mathcal{L}_{M}$ describes the matter and 
 additional couplings with the field $B_{\mu}$. The potential $V$ is responsible for triggering the 
 spontaneous Lorentz violation in case of the bumblebee field assumes a nonzero
  VEV $\left\langle B_{\mu}\right\rangle \equiv b_{\mu}$, 
satisfying the condition $B^{\mu}B_{\mu}=\mp b^{2}$. It is worth emphasizing that the quantity $b^{2}$ 
is a positive real number, and the $\pm$ sign implies that $b_{\mu}$ is timelike or spacelike, respectively.
 The model described by the action (\ref{S1}) and other versions involving different couplings or choices of the potential 
have been investigated in a variety of contexts (as mentioned in Introduction).

The gravitational field equations in the bumblebee context or gravity can be directly obtained  by varying the 
action (\ref{S1}) with respect to
the metric tensor $g_{\mu\nu}$, while keeping the bumblebee field $B_{\mu}$ fixed. That procedure yields 
\begin{eqnarray}
&& G_{\mu\nu}+\Lambda g_{\mu\nu}= \kappa \left( T^{B}_{\mu\nu}+ T^{M}_{\mu\nu} \right) \nonumber \\
	&&= \kappa\left[2V'B_{\mu}B_{\nu} +B_{\mu}^{\ \alpha}B_{\nu\alpha}-\left(V+ \frac{1}{4}B_{\alpha\beta}B^{\alpha\beta}\right)g_{\mu\nu} \right] \nonumber\\
	&& +\xi\left[\frac{1}{2}B^{\alpha}B^{\beta}R_{\alpha\beta}g_{\mu\nu}-B_{\mu}B^{\alpha}R_{\alpha\nu}-B_{\nu}B^{\alpha}R_{\alpha\mu}\right.\nonumber\\
	&& +\frac{1}{2}\nabla_{\alpha}\nabla_{\mu}\left(B^{\alpha}B_{\nu}\right)+\frac{1}{2}\nabla_{\alpha}\nabla_{\nu}\left(B^{\alpha}B_{\mu}\right) \nonumber\\
	&& \left.-\frac{1}{2}\nabla^{2}\left(B_{\mu}B_{\nu}\right)-\frac{1}{2}
g_{\mu\nu}\nabla_{\alpha}\nabla_{\beta}\left(B^{\alpha}B^{\beta}\right)\right] +\kappa T^{M}_{\mu\nu},
\label{modified}
\end{eqnarray}
as  modified gravitational field equations, in which 
$G_{\mu\nu}$ is the Einstein tensor and the operator $'$ means derivative with respect to the potential argument.
 In the general case,
$T^{B}_{\mu\nu}$ and $T^{M}_{\mu\nu}$ are the energy-momentum tensors of the bumblebee field and of the matter field, respectively. In order to solve Eq. (\ref{modified}), it is necessary to choose a bumblebee potential $V$
 and a metric \textit{Ansatz} with some symmetry, like the spherical or the axial symmetry. By doing that, one
 obtains a set of equations and, solving them,  a full metric. Here we focus on the spherical symmetry and comment
 some potentials as options to get a full black hole metric, whether with or without a cosmological constant.
 
 The action (\ref{S1}) also provides an equation of motion for $B^{\mu}$. By varying that action in this time with
 respect to the bumblebee field leads to
 \begin{equation}
 \nabla_{\mu}B^{\mu\nu}=2\left( V'B^{\nu}-\frac{\xi}{2\kappa}B_\mu R^{\mu\nu}  \right).
 \label{B_eq}
 \end{equation}
With all framework introduced, in which we do not consider a coupling between the bumblebee field and
the matter field, we will apply it to the nonzero cosmological constant case, generating then
black holes with a cosmological constant. But before that, we comment a previous result that involves a
null cosmological constant. 

\subsection{The $\Lambda=0$ case}

In this framework, an exact black hole solution without a cosmological constant  was constructed by Casana \textit{et al.}
 \cite{Casana2018}, also known as the Schwarzschild-like black hole. According to the authors, 
 a spherically symmetric  spacetime  in the absence of both matter
  $(\mathcal{L}_{M} = 0)$ and a cosmological constant ($\Lambda=0$) was interpreted as
  a Schwarzschild-like black in the bumblebee gravity, from a radial bumblebee field $B_{\mu}$ written as
\begin{equation}
    B_\mu=b_{\mu}=(0,b_{r}(r),0,0).
    \label{VEV}
\end{equation}
In the coordinates $(t,r,\theta,\phi)$, the general spherical  \textit{Ansatz} used by the authors (and adopted here)
 is given by
\begin{equation}
g_{\mu\nu}=\mbox{diag}(-e^{2\gamma(r)}, e^{2\rho(r)},r^2, r^2\sin^2\theta).
\label{Ansatz}
\end{equation}
Adopting both the mentioned \textit{Ansatz} and the condition $b_{\mu}b^{\mu}=b^2=\mbox{const.}$, one has
the radial component of the bumblebee field when it assumes the VEV, i.e., 
\begin{equation}
b_{r}(r) =\left\vert b\right\vert e^{\rho (r)}.
\label{bb}
\end{equation}
As we can see, contrary to Bertolami and Páramos \cite{Bertolami2005}, the authors of Ref.  \cite{Casana2018}, 
from  Eq. (\ref{bb}), have $\nabla_{\mu}b_{\nu}\neq 0$, which is the same form of the bumblebee field that 
we will adopt next.  It is worth pointing out that the vanishing condition for the covariant derivative is just 
possible for special spacetimes with a geometrical constraint that comes from the underlying
pseudo-Riemann geometry assumption.

With that \textit{Ansatz} plus the mentioned bumblebee form and assuming $V=V'=0$, the following metric
\begin{eqnarray}
 ds^{2}&=&-\left(1-\frac{2M}{r}\right)dt^{2}+(1+\ell) \left(1-\frac{2M}{r}\right)^{-1}dr^{2}\nonumber \\
 && +r^2 \left(d\theta^2+ \sin^2\theta d\phi^2 \right)
  \label{Casana}
\end{eqnarray}
is a solution of the modified field equations (\ref{modified}). The metric (\ref{Casana}) is also called 
Schwarzschild-like geometry. The bumblebee field or the Lorentz-violating parameter is represented in that metric
 by $\ell=\xi b^2$, and the parameter $M$ stands
for the usual mass of the Schwarzschild black hole in the limit $\ell\rightarrow0$. Note that the 
condition (\ref{VEV}) and the potential choice imply that the field $B_{\mu}$ stays frozen in its VEV $b_{\mu}$,
 that is to say,  $V = 0$ (vacuum condition) and the assumption $V' = 0$ ensures that the field is in the minimum of the potential. Besides that, since the background field $b_{\mu}$ is a spacelike vector purely radial, its associated field strength is identically null, i.e., $B_{\mu\nu}=0$. 
 
  A solution like (\ref{Casana}) resembles that one obtained by Seifert in Ref. \cite{Seifert}, 
 in which a Lorentz-violating topological defect was studied. In the mentioned paper, 
 the author presented a  topological defect
  solution from a Lorentz symmetry breaking triggered spontaneously by a tensor field, namely,
  a rank-two antisymmetric tensor field. Such a solution was interpreted as a vacuum monopole 
  solution. As we said, such a monopole solution reseambles (\ref{Casana}), but it does not approach 
  asymptotically the line element (\ref{Casana}) due to a different $r$-dependence. On the other hand, according to
  Ref. \cite{Seifert2}, for the Lorentz symmetry breaking triggered by a vector field, 
  which is just the case considered here, a domain wall
  (another kind of topological defect) solution was obtained only when a timelike vector
  was adopted. In this sense, the possibility of a topological defect solution for a spacelike vector (like the
  bumblebee field adopted here) is prohibited.

From the metric (\ref{Casana}), it is clear to see that the event horizon does not depend on $\ell$. As is well known,
considering a metric like (\ref{Ansatz}), zeros of $g^{rr}=e^{-2\rho(r)}=0$ provide the localization of horizons.
As we can directly read, for the metric (\ref{Casana}) one has $r_+=2M$, the same value of the event
horizon of the Schwarzschild black hole. In the same way, as we will see, the photon sphere is located at $r_{ph}=3M$,
like Schwarzschild's. Such a surface is responsible for the black hole shadow. On the other hand, as
 pointed out by Casana \textit{et al.} \cite{Casana2018}, the light
bending and the perihelion advance bring out the Lorentz-violating parameter and its (possible) tiny 
influence.\footnote{In Ref. \cite{Casana2018}, there are upper bounds on the parameter $\ell$ from, 
for example, light deflection, time delay of light, and perihelion advance of the planet Mercury. 
The most stringent upper bound on the Lorentz-violating parameter is $\ell < 10^{-15}$ to date.}
 It is worth emphasizing that the solution (\ref{Casana}) cannot be converted into the standard 
 Schwarzschild solution for a nonzero value of $\ell$ by means a suitable coordinate 
 transformation \cite{Casana2018,Debora2020}. In this sense, the metric (\ref{Casana}) is an entirely 
 new spacetime metric.

\subsection{The $\Lambda \neq 0$ case}

Following the approach outlined above, we will now investigate some effects of the Lorentz violation in the presence of a 
nonzero cosmological constant on the model described by the action (\ref{S1}). More specifically, we are 
interested in obtaining an exact black hole solution in the presence of a cosmological constant, a geometry similar to 
either the Schwarzschild-anti-de Sitter black hole or the Schwarzschild-de Sitter black hole (depending on the sign of the
cosmological constant). One route to be explored here is to relax the vacuum conditions, i.e., $V = 0$
 and $V' = 0$ assumed by Casana \textit{et al.} \cite{Casana2018}. A simple example of  a potential that 
 satisfies such conditions is clearly provided by a smooth quadratic form
\begin{equation}
V=V(X)=\frac{\lambda}{2}X^{2},
\end{equation}
 where $\lambda$ is a constant, and $X$ is a generic potential argument. In this case, the VEV $b_{\mu}$ is 
 solution of $V=V'=0$. Another simple choice of potential consists of a linear function 
\begin{equation}\label{potlinear}
V=V(\lambda,X) = \frac{\lambda}{2}X,
\end{equation}
 where now $\lambda$ is a Lagrange-multiplier field \cite{Bluhm2008}. Note that the equation of motion 
 for the Lagrange-multiplier ensures the vacuum condition $X = 0$, and then  $V=0$ for any field
  $\lambda$ on-shell. However, for the linear functional form (\ref{potlinear}), it follows that $V'\neq 0$ when the 
  field $\lambda$ is nonzero, so additional contributions from the potential $V$ can  modify the Einstein equations.
   Since $\lambda$ has no kinetic terms, it is auxiliary and cannot propagate. 
   However, it is also an additional degree of freedom that appears in the equations of motion. In fact, 
   the equations of motion for the metric (\ref{modified}) and the bumblebee field (\ref{B_eq}) provide 
   constraints on the field $\lambda$. Moreover, in order to be well defined, all bumblebee models require explicit initial 
   conditions on the field excitations about vacuum values. Like the bumblebee field, the Lagrange multiplier
    can also be expanded around its vacuum value $\left\langle \lambda \right\rangle$ as
 \begin{equation}
 \lambda=\left\langle \lambda \right\rangle+\tilde{\lambda}.
 \end{equation}
For our purpose here, it is convenient to fix the initial conditions taking $\tilde{\lambda}=0$, which implies that
 the field $\lambda$  remains frozen in its VEV. This is a similar hypothesis used 
 for the bumblebee field in Ref. \cite{Casana2018}. \textit{A priori}, $\left\langle \lambda \right\rangle$ could 
 depend on the spacetime position, but it is sufficient to assume it as a real constant. 
 Thus, in what follows, the on-shell value of $\lambda$ is given by
  $\lambda\equiv \left\langle \lambda \right\rangle$, with its value fixed from the equations 
  of motion (\ref{modified}) or (\ref{B_eq}) in terms of other parameters of the model.

It is worth noting that the action (\ref{S1}) with the conditions $V=V'=0$, adopted by Casana \textit{et al.}
 \cite{Casana2018}, will provide a black hole solution from the \textit{Ansatz} (\ref{Ansatz}) only if $\Lambda=0$.
 A black hole solution with a nonzero cosmological constant needs a different potential, in the case of
 a solution with the relation $ e^{2\gamma(r)}=(1+\ell)e^{-2\rho(r)}$ exhibited in the Schwarzschild-like
 black hole (\ref{Casana}).
 Therefore, in order to build black holes with a cosmological constant, we assume, for that purpose, the linear potential
 written as
 \begin{equation}
V(B^{\mu}B_{\mu}- b^{2}) = \frac{\lambda}{2}(B^{\mu}B_{\mu}- b^{2}) \ \    \mbox{and} \ \   V'=  \frac{\lambda}{2},
\label{Potential}
\end{equation}
where $B_\mu$ is assumed radial-like (\ref{VEV}). With such a potential and the \textit{Ansatz} (\ref{Ansatz}), one has three independent equations 
from Eq. (\ref{modified}) because $G_{\phi\phi}=\sin^2\theta G_{\theta\theta}$.
 That is to say, our system of equations reads
 \begin{equation}
 \rho'(r) -\frac{1}{2r}\left[ 1-\frac{\left( 1-\Lambda r^2 \right)}{\left(1+\ell \right)}e^{2\rho(r)}  \right]=0,
 \label{System1}
 \end{equation}
\begin{eqnarray}
&&\gamma''(r)+\gamma'(r)^2-\frac{2}{r}\left[ \frac{(1+\ell)}{\ell}\gamma'(r)+\rho'(r) \right]-\gamma'(r)\rho'(r) \nonumber \\
&&-\frac{1}{\ell r^2}\left[(1+\ell )-\left(1+(\kappa\lambda b^2-\Lambda)r^2 \right)e^{2\rho(r)} \right]=0, \label{System2} \\
&&\gamma''(r)+\gamma'(r)^2+\frac{1}{r}\left[ \gamma'(r)-\rho'(r) \right] -\gamma'(r)\rho'(r) \nonumber \\
&&+\frac{\Lambda}{(1+\ell)}e^{2\rho (r)}=0.
\label{System3}  
\end{eqnarray}
 As we can see, our system shows three independent equations and two unknown functions ($\gamma$ and $\rho$).
  A third would come from the matter energy-momentum tensor with a suitable equation of state. 
  In our case, without a matter field,  that function is zero. 
 
 Equation (\ref{System1}) is a differential equation which involves just $\rho (r)$. Thus, its solution is directly
given by
\begin{equation}
\rho (r)=\frac{1}{2}\ln \left[(1+\ell)\left(1-\frac{C_1}{r}-\frac{\Lambda}{3}r^2 \right)^{-1}  \right],
\label{rho}
\end{equation}    
where $C_1$ is an integration constant interpreted as some sort of mass parameter of the Schwarzschild geometry. By making $\Lambda=0$, we hope to recover the Schwarzschild-like solution. In this sense, $C_1 = 2M$ for that purpose.
 
In order to generate a solution of Eqs. (\ref{System1})-(\ref{System3}) similar to the $\Lambda=0$ case, we
use the mentioned relation between the metric terms, $ e^{2\gamma(r)}=(1+\ell)e^{-2\rho(r)}$. Thus from 
Eq. (\ref{rho}), one has  
\begin{equation}
\gamma (r)=\frac{1}{2}\ln \left[1-\frac{2M}{r}-\frac{\Lambda}{3}r^2  \right]
\label{gamma}
\end{equation}    
as solution of our system. Such a relation between the metric terms provides $\gamma (r)$ and an appropriate solution 
for the system of equations
  (\ref{System1})-(\ref{System3}) even with a nonzero cosmological constant.
However, with the potential given by (\ref{Potential}), a solution of this type will be possible if and only if
\begin{equation}
\Lambda=\frac{\kappa\lambda}{\xi}(1+\ell).
\label{Constraint}
\end{equation}
The constraint (\ref{Constraint}) is a \textit{conditio sine qua non} in order to generate a metric 
with a cosmological constant from the modified Einstein equations (\ref{modified}) and the potential (\ref{Potential}).
This is the constraint on the field $\lambda$ coming from the modified Einstein field equations 
as mentioned earlier.

With all metric terms known, namely  (\ref{rho}) and (\ref{gamma}), our proposed metric with spherical symmetry and a
cosmological constant reads 
\begin{widetext} 
\begin{equation}
 ds^{2}=-\left(1-\frac{2M}{r}-(1+\ell)\frac{\Lambda_e}{3}r^2\right)dt^{2}+(1+\ell) \left(1-\frac{2M}{r}-(1+\ell)\frac{\Lambda_e}{3}r^2\right)^{-1}dr^{2} +r^2 \left(d\theta^2+ \sin^2\theta d\phi^2 \right),
  \label{Metric}
\end{equation}
\end{widetext} 
in which, by convenience, we conceive of $\Lambda_{e}=\frac{\kappa\lambda}{\xi}$ as an effective cosmological
constant. It is worth mentioning that the constraint (\ref{Constraint}) also guarantees the energy conservation 
of the bumblebee energy-momentum tensor. Indeed, 
\begin{equation}
\nabla _{\nu}T_B^{\mu\nu}=0
\end{equation}
for all components, except for the $T_B^{rr}$ component that asks that constraint in order to satisfy
 the energy-momentum conservation. And the equation of motion for the bumblebee field (\ref{B_eq})
  is also verified by using the constrain (\ref{Constraint}) for the nonzero cosmological 
  constant case.

The metric (\ref{Metric}) is richer than the metric (\ref{Casana}) that excludes a cosmological constant. 
As will see, the bumblebee field influence is found even on the horizons, contrary to the Schwarzschild-like
solution in which the event horizon radius is the same of the Schwarzschild black hole. In particular, the last part of 
this article will show the influence of the bumblebee field on the shadow angular radius. Those influences are a 
consequence of the Lorentz symmetry violation, which is translated into geometry or into the general relativity 
language from a privileged spacetime direction or an anisotropic fluid. This is noteworthy from the 
total energy-momentum tensor of the metric (\ref{Metric}) given by
\begin{equation}
T_{\nu}^{\mu}=\frac{1}{\kappa}\left(\begin{array}{cccc}
-\epsilon\\
 &p_r \\
 &  & p_t \\
 &  &  & p_t
\end{array}\right),
\label{Energy-momentum}
\end{equation}
in which
\begin{equation}
p_r=-\epsilon=-\frac{\ell}{(1+\ell)r^2}-\Lambda_e \ \ \ \mbox{and} \ \ \ p_t=-\Lambda_e,
\end{equation}
 with $\epsilon$ and $p_r$ playing the role of the 
energy density and the radial pressure, respectively, and $p_t$ is the tangential pressure. 
Then the anisotropic feature of the spacetime (\ref{Metric})
gets evident, the radial and tangential pressures are different. As we can see, in particular, the radial pressure
 is always negative  when $\Lambda_e >0$, which is the Schwarzschild-de Sitter-like case as we will
  indicate later.
A relation between the bumblebee field (specifically its potential) and a de Sitter phase in
cosmology was already indicated in Ref. \cite{cosmology} and it appears in the gravitational context once again here. 

It is worth emphasizing that the metric (\ref{Metric}) is neither asymptotically de Sitter nor asymptotically 
anti-de Sitter. That is, we cannot write that metric in a particular form such that, in the end of the day, 
\begin{equation}
\lim_{r \rightarrow \infty} g_{tt}= \lim_{r \rightarrow \infty} g^{rr}= \lim_{r \rightarrow \infty}1-\frac{\Lambda}{3}r^2.
\end{equation}
The factor before $g_{rr}$, namely $(1+\ell)$, forbids the above limit. According to Ref. \cite{Debora2020}
mentioned before, that factor also
forbids a coordinate transformation that turns the Schwarzschild-like black hole into the Schwarzschild black
hole. The same argument can be adopted here, for our metric uses the $e^{2\gamma(r)}=(1+\ell)e^{-2\rho(r)}$
relation such as the Schwarzschild-like black hole. Therefore, the metric (\ref{Metric}) is not converted into either
the Schwarzschild-anti-de Sitter black hole or the Schwarzschild-de Sitter black hole.

Another feature of the metric (\ref{Metric}) regards to real or \enquote{fictitious} singular points. The former is the
physical singularity that appears by taking, for example, the limit of the Kretschmann scalar $K$ (built from the Riemann tensor), which is given by
\begin{eqnarray}
K&=&R_{\alpha\beta\delta\gamma} R^{\alpha\beta\delta\gamma}=\frac{8}{3}\Lambda_e^2 +\frac{\ell}{(1+\ell)r^2} \bigg [ \frac{8}{3}\Lambda_e+\frac{4\ell}{(1+\ell)r^2} \nonumber \\
&&+ \frac{16M}{(1+\ell)r^3}+\frac{48M^2}{\ell (1+\ell)r^4} \bigg ].
\end{eqnarray}
The Schwarzschild-like singularity is present only at $r=0$ (considering $0<\ell \ll 1$).
 The mentioned \enquote{fictitious} singular points are, 
indeed, a bad choice of the coordinate system. As we said, zeros of $g^{rr}$ give us not a \enquote{physical singularity},
but special surfaces, horizons, which depend on the sign of $\Lambda_e$ in the metric (\ref{Metric}).

 The metric (\ref{Metric}) 
is also independent of two coordinates: $t$ and $\phi$. Thus, that geometry possesses two Killing vector 
fields ($\partial/ \partial t$ and $\partial/ \partial \phi$) related to two conserved quantities, 
energy and angular momentum (both will be useful later for the geodesic motion).

\subsubsection*{Schwarzschild-anti-de Sitter-like black hole}
\begin{figure}
\begin{center}
\includegraphics[scale=0.75]{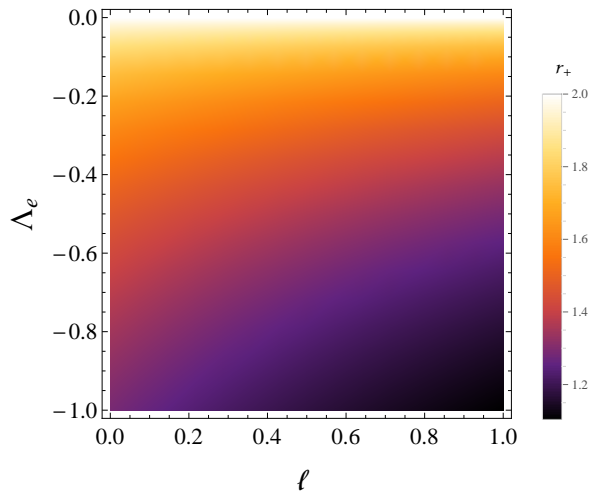}
\caption{Relations between the metric parameters and the event horizon radius, $r_+$,
for the Schwarzschild-anti-de Sitter-like case. $\Lambda_e$ and $\ell$ stand for the 
 effective cosmological constant and the Lorentz-violating parameter, respectively. 
 As we can see, that parameter can decrease the event horizon radius. In this graphic, we adopted $M=1$.}
\label{Ads}
\end{center}
\end{figure}

Assuming $\Lambda_e <0$, we have a Schwarzschild-anti-de Sitter-like solution. And its spacetime structure
 is given by a unique horizon, the event horizon, whose radius is
\begin{equation}
r_+=-\frac{1}{\mathcal{F}(M,\Lambda_e,\ell)^{\frac{1}{3}}} - \frac{\mathcal{F}(M,\Lambda_e,\ell)^{\frac{1}{3}}}{(1+\ell)\Lambda_e},
\end{equation} 
with
\begin{equation}
\mathcal{F}(M,\Lambda_e,\ell)=(1+\ell)^2 \Lambda_e^2 \left(3M+\sqrt{9M^2-\frac{1}{(1+\ell)\Lambda_e}}  \right) .
\end{equation}
The relation between the parameters of the black hole (\ref{Metric}) and the event horizon radius is indicated 
in Fig. \ref{Ads}.
As we can see, for large values of $\vert \Lambda_e \vert$, the symmetry breaking parameter, $\ell$, 
decreases the horizon radius.

In this case, $\Lambda_e < 0$, the unique Killing surface coincides with the event horizon. The Killing surfaces
localization is calculated from $g_{tt}=0$, they are surfaces where the Killing vector field $\partial / \partial t$ is
null or lightlike. Above all, from that surface to infinity,
the Killing vector field $\partial / \partial t$ is timelike in the Schwarzschild-anti-de Sitter-like case. 
In this entire region, $r>r_+$,  static observers are viable ones.

\subsubsection*{Schwarzschild-de Sitter-like black hole}
\begin{figure}
\begin{center}
\includegraphics[scale=0.75]{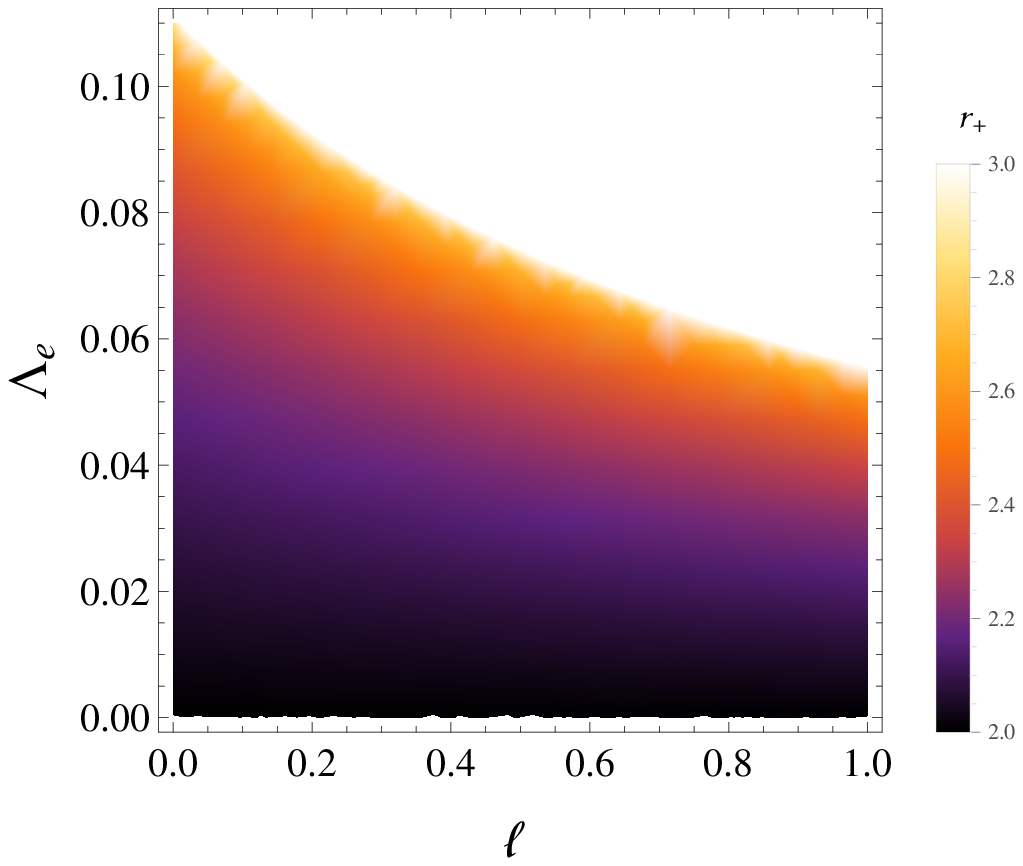}
\includegraphics[scale=0.75]{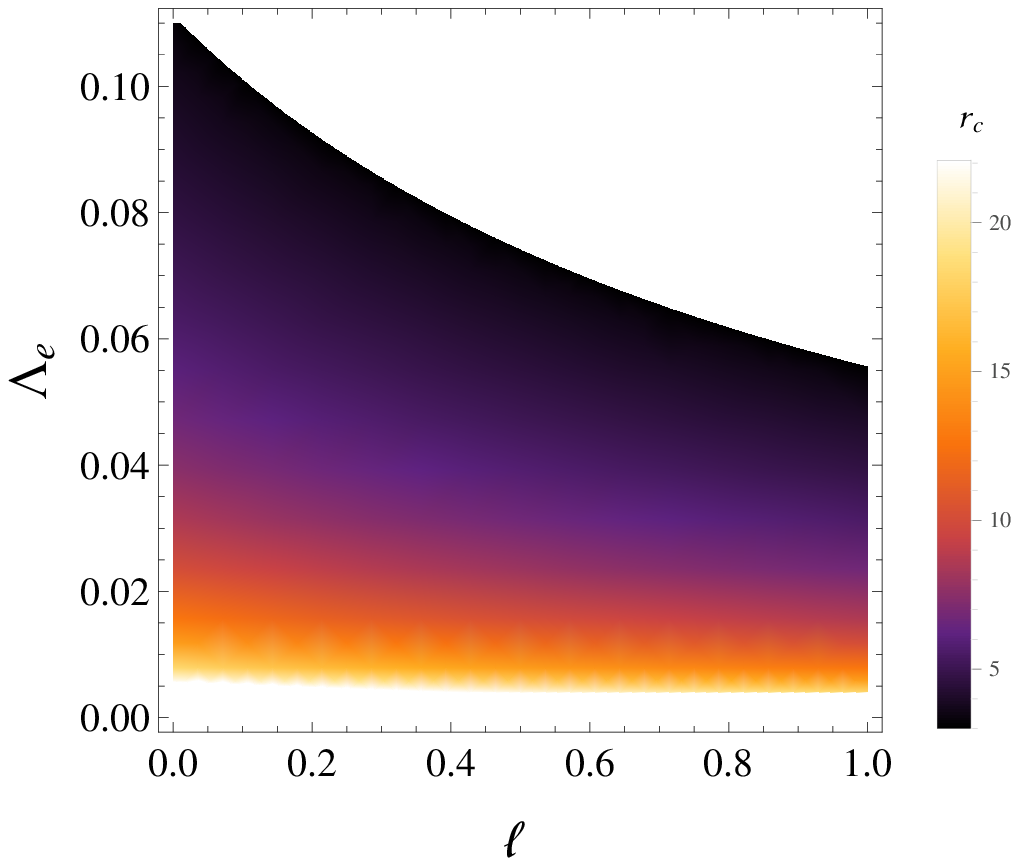}
\caption{Relations between the metric parameters and the event horizon radius, $r_+$, and the cosmological
 horizon radius, $r_c$, for the Schwarzschild-de Sitter-like case. $\Lambda_e$ and $\ell$ stand for the 
 effective cosmological constant and the Lorentz-violating parameter, respectively. 
 The parameter $\ell$ may increase the event horizon and decrease the cosmological horizon.
  In this graphic, we adopted $M=1$, thus white regions are forbidden for that value.}
\label{dS}
\end{center}
\end{figure}

On the other hand, the Schwarzschild-de Sitter-like black hole appears from $\Lambda_e >0$. In this case, the spacetime
structure presents two horizons: the event horizon, $r_+$, and the cosmological horizon, $r_c$. However, in order
to provide two horizons (two real roots of $g^{rr}=0$), the metric parameters or the effective cosmological constant should obey 
the following inequality:
\begin{equation}
0<\Lambda_e < \frac{1}{9M^2(1+\ell)}.
\end{equation} 
With $\ell =0$ one has the well-known Schwarzschild-de Sitter condition to generate two horizons.\footnote{For
a review on the Schwarzschild-anti-de Sitter and Schw- \ arzschild-de Sitter geometries, see Ref. \cite{Stuchlik}.} And
for $9M^2(1+\ell)\Lambda_e =1$, our metric shows an extreme case, in which $r_+=r_c$.  Analytic expressions
for the event and the cosmological horizons are, respectively,  
\begin{equation}
r_+ = \frac{2}{\sqrt{(1+\ell)\Lambda_e}}\cos \left(\frac{\pi}{3}+\frac{\alpha}{3} \right),
\end{equation}
\begin{equation}
r_c =\frac{2}{\sqrt{(1+\ell)\Lambda_e}}\cos \left(\frac{\pi}{3}-\frac{\alpha}{3} \right),
\end{equation}
with $\alpha=\cos^{-1} (3M\sqrt{(1+\ell)\Lambda_e})$. As we can see in Fig. \ref{dS}, the 
Lorentz-violation parameter, $\ell$, modifies the spacetime structure, it decreases the cosmological 
horizon radius and, at the same time, increases
the event horizon radius for considerable values of $\Lambda_e$.  The region $r_+<r<r_c$ is the so-called domain
 of outer communication. In that region,
 observers  may  communicate with each other, thus there is no horizon in that region. 
 But an observer beyond the cosmological horizon may be invisible according to someone inside the
 domain of outer communication. In the domain of outer communication, observers can be static ones, and
  the Killing vector field $\partial /\partial t$ is timelike in that region (consequently, having both the spherical
  symmetry and the mentioned timelike Killing vector makes Eq. (\ref{Metric}) a static spacetime).
 In the next section, we calculate the shadow angular radius of the metric (\ref{Metric}) as seen by a static observer 
 in the domain of outer communication.
 
\section{The shadow angular radius}
The black hole shadow is a dark region in the bright sky caused by a black hole and its huge gravitational field or
the intense light deflection. 
Here we are interested in calculating the angular radius of
the shadow generated by the black hole (\ref{Metric}) for $\Lambda_e >0$, the Schwarzschild-de Sitter-like case
(more appropriate from the cosmological point of view). 
As we mentioned, our observer will be at rest in the domain of outer communication, that is to say, our observer
is a static one. The shadow silhouette is given by unstable orbits (circular unstable orbits for our metric) outside
the event horizon. In such orbits, photons may either go into the black hole or go to the opposite direction, reaching,
for example, our observer. Therefore, we need the null geodesic equations in order to obtain such special 
orbits and trace them to the observer position.

\begin{figure}
\begin{center}
\includegraphics[trim=0.5cm 0.2cm 0cm 0cm, clip=true,scale=0.52]{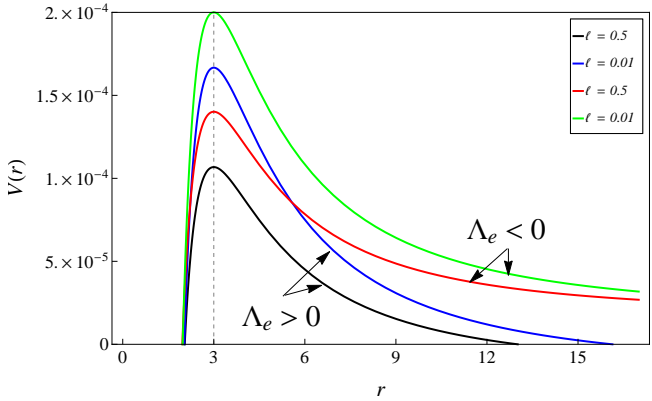}
\caption{Gravitational potential (\ref{Grav_Potential}) and the influence of the Lorentz-violating parameter $\ell$ on it.
As we can see, the parameter $\ell$ does not modify the photon sphere radius, indicated by $r_{ph}=3M$.
In this graphic one adopts $M=1$ and $\vert \Lambda_e \vert$=0.01. }
\label{V}
\end{center}
\end{figure}

Geodesics are calculated from the Lagrangian
\begin{equation}
\mathcal{L}=\frac{1}{2}g_{\mu\nu}\dot{x}^{\mu} \dot{x}^{\nu},
\end{equation}
where dot means derivative with respect to the affine parameter of the curve (indicate here by $\tau$).
 In particular, an equatorial null geodesic
($\theta=\pi/2$) for  the metric (\ref{Metric}) becomes simply
\begin{eqnarray}
&&-\left(1-\frac{2M}{r}-(1+\ell)\frac{\Lambda_e}{3}r^2\right)\dot{t}^2 \nonumber \\
&&+(1+\ell)\left(1-\frac{2M}{r}-(1+\ell)\frac{\Lambda_e}{3}r^2\right)^{-1}\dot{r}^2+r^2\dot{\phi}^2=0.\nonumber \\
\label{Geodesics}
\end{eqnarray}
As we said, our proposed metric has $\partial /\partial t$ and $\partial / \partial \phi$ as two Killing 
vector fields that, consequently, yield two conserved quantities.
 The first one ($\partial /\partial t$) provides the energy conservation, and the second one  ($\partial / \partial \phi$)
  gives us the angular momentum conservation. Therefore, photons along the geodesics (\ref{Geodesics}) 
  have both conserved energy $E$ and angular momentum $L$ given by
\begin{equation}
E=\left(1-\frac{2M}{r}-(1+\ell)\frac{\Lambda_e}{3}r^2\right)\dot{t} \ \ \  \mbox{and} \ \ \ L=r^2\dot{\phi}.
\end{equation}
As is known from textbooks, the geodesic equations from a spacetime with spherical symmetry like (\ref{Metric}) 
may be written as a conservation equation for each photon, a familiar equation like
\begin{equation}
\frac{1}{2}\left(\frac{dr}{d\tau} \right)^2+\mathcal{V}(r)=\mathcal{E},
\end{equation} 
in which the gravitational potential $\mathcal{V}(r)$ and the energy $\mathcal{E}$ are defined as
\begin{equation}
\mathcal{V}(r)=\frac{L^2}{2(1+\ell)} \left[\frac{1}{r^2}-\frac{2M}{r^3}-(1+\ell)\frac{\Lambda_e}{3}   \right],
\label{Grav_Potential}
\end{equation}
\begin{equation}
\mathcal{E}=\frac{E^2}{2(1+\ell)}.
\end{equation}
The condition of unstable orbits ($d\mathcal{V}(r)/dr=0$ and $d^2\mathcal{V}(r)/dr^2<0$) that 
compose the black hole shadow give us the following
radius: $r_{ph}=3M$. This is the photon sphere radius, and such a sphere is the surface that creates 
the shadow silhouette. 
Therefore, in order to measure the shadow
angular radius, our observer will be beyond the photon sphere. As we can see, 
that radius is the same of the Schwarzschild photon sphere, as much as of the Schwarzschild-de Sitter sphere.
The Lorentz-violating parameter $\ell$ does not modify such a result (see Fig. \ref{V}). 

Circular orbits implies $dr/d\phi=0$, thus Eq. (\ref{Geodesics}) delivers a useful ratio
\begin{equation}
\frac{E^2}{L^2}=\frac{1}{r^2}-\frac{2M}{r^3}-(1+\ell)\frac{\Lambda_e}{3}.
\end{equation}
With the aid of the photon sphere radius, $r_{ph}=3M$, photons that compose the shadow silhouette 
present the following constant energy-angular momentum ratio:
\begin{equation}
\frac{E^2}{L^2}=\frac{1}{27M^2}-(1+\ell)\frac{\Lambda_e}{3}.
\label{ratio}
\end{equation}

\begin{figure}
\begin{center}
\includegraphics[scale=0.7]{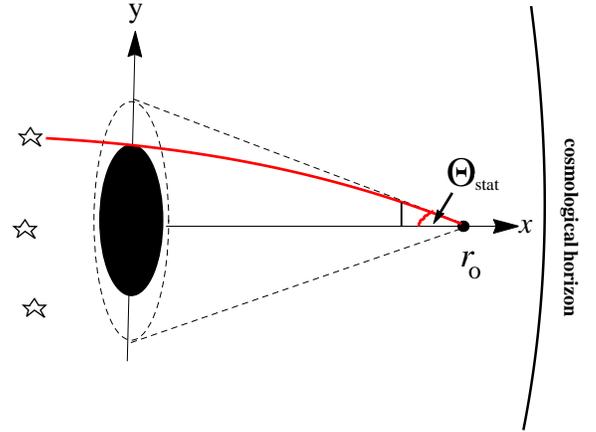}
\caption{Definition of the shadow angular radius $\Theta_{\mbox{stat}}$ for a static observer at rest ($r= r_o$).}
\label{Draft}
\end{center}
\end{figure}

We are going to follow the approach presented in Ref. \cite{Perlick} where  the shadow angular radius
was calculated for the Schwarzschild-de Sitter black hole. In that approach, our static
observer---as we said---is in the domain of outer communication, between the photon sphere and
the cosmological horizon. His/her radial coordinate is $r=r_o$. According to Fig \ref{Draft},
 the shadow angular radius is defined as
\begin{equation}
\tan \Theta=\lim_{\Delta x \rightarrow 0} \frac{\Delta y}{\Delta x}.
\end{equation}
In order to obtain the angle $\Theta$, we adopt the isotropic coordinate system, in which angles are invariant in
comparison with the Euclidian space. With that coordinate system, our two-dimensional metric 
(spatial components with $\theta=\pi/2$) becomes conformal to the two-dimensional Euclidian space in spherical coordinates and can be written as
\begin{equation}
ds_{2}^2=\Omega(\bar{r})^2\left(d\bar{r}^2+\bar{r}^2d\phi^2 \right),
\end{equation}
from the following  transformations
\begin{equation}
r^2=\Omega(\bar{r})^2 \bar{r}^2,
\end{equation}
\begin{equation}
(1+\ell) \left(1-\frac{2M}{r}-(1+\ell)\frac{\Lambda_e}{3}r^2\right)^{-1}dr^{2}=\Omega(\bar{r})^2 d \bar{r}^2,
\end{equation}
in which $\Omega (\bar{r})$ is the conformal factor. Therefore, from the above relations, it is easy to get
\begin{eqnarray}
\left(\frac{dy}{dx} \right)^2 & =& \bar{r}^2 \left( \frac{d\phi}{d\bar{r}}\right)^2=\frac{r^2}{(1+\ell)} \nonumber \\
&\times&  \left(1-\frac{2M}{r}-(1+\ell)\frac{\Lambda_e}{3}r^2\right) \left( \frac{d\phi}{dr}\right)^2.
\end{eqnarray}
The term $d\phi/dr$ comes directly from the geodesic equation (\ref{Geodesics}), and with a useful 
trigonometric relation, namely $\sin^2 \Theta=\tan^2 \Theta/(1+\tan^2 \Theta)$, the following angle is obtained
\begin{equation}
\sin^2 \Theta=\left(1-\frac{2M}{r}-(1+\ell)\frac{\Lambda_e}{3}r^2\right) \left(\frac{E}{L}r \right)^{-2}.
\label{Sin}
\end{equation}
For a static observer at $r=r_o$ in the domain of outer communication, the shadow angular radius is obtained from
orbits with the calculated energy-angular momentum ratio (\ref{ratio}), that is to say, orbits from the photon sphere.
 Then substituting Eq. (\ref{ratio}) into Eq. (\ref{Sin}),
 we have the sought-after relation
 \begin{equation}
  \sin^2 \Theta_{\mbox{stat}}=\frac{\left(1-\frac{2M}{r_o}-(1+\ell)\frac{\Lambda_e}{3}r_o^2\right)}{\left( \frac{1}{27M^2}-(1+\ell)\frac{\Lambda_e}{3}\right)r_o^2}. 
  \label{Angular_radius}
 \end{equation}
As we can see, the above relation equals the result obtained in Ref. \cite{Perlick} by imposing $\ell=0$, which is the 
Schwarzschild-de Sitter black hole studied in the mentioned article. Moreover, with $\ell=\Lambda_e=0$, we recover
the shadow angular radius of the Schwarzschild geometry. As we mentioned earlier, a geometry like (\ref{Metric}),
with $ e^{2\gamma(r)}=(1+\ell)e^{-2\rho(r)}$, cannot be converted into either the usual Schwarzschild solution
($\Lambda_e = 0$), or the Schwarzschild-anti-de Sitter solution ($\Lambda_e < 0$), or the Schwarzschild-de Sitter solution ($\Lambda_e > 0$) from a suitable coordinate transformation. Therefore, the parameter $1+ \ell$ cannot be
absorbed into the  cosmological constant, and then the result (\ref{Angular_radius}) does not turn into the shadow 
angular radius of the Schwarzschild-de Sitter black hole as calculated in Ref. \cite{Perlick}.

\begin{figure}
\begin{center}
\includegraphics[trim=0cm 0.2cm 1.5cm 0cm, clip=true,scale=0.68]{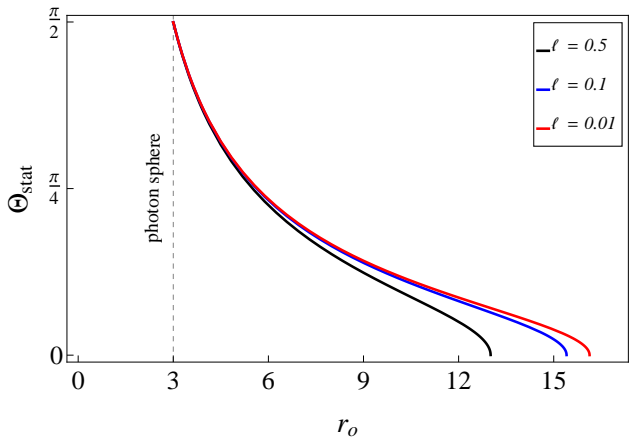}
\caption{Shadow angular radius $\Theta_{\mbox{stat}}$ as seen by a static observer at different locations ($r=r_o$). 
The dashed line indicates the photon sphere, $r_{ph}=3M$, which is the surface that causes the black hole shadow. 
As we can see, the larger the Lorentz-violating parameter $\ell$ is, less the shadow angular radius. If the angular radius
is zero, our observer will be at the cosmological horizon.  
In this graphic, we adopted $M=1$ and $\Lambda_e = 0.01$.}
\label{Radius}
\end{center}
\end{figure}

Some highlights of the result (\ref{Angular_radius}): for $r_o=r_c$, the shadow angular radius is zero, i.e., the
sky is entirely bright for our observer. On the other hand, for $r_o=r_{ph}$, the shadow angular radius is maximum, thus
observer's sky is half bright and half dark. Lastly, according to Fig. \ref{Radius}, we can see the influence of the 
Lorentz-violating parameter on the shadow. The influence of the parameter $\ell$ on the shadow was commented in
Ref. \cite{Casana2020} where a Kerr-like black hole was built in the bumblebee gravity. 
That parameter increases the shadow deformation
 when a rotating black hole is considered. But, for the first time in the literature, the influence of the 
Lorentz-violating parameter on the shadow angular radius was straightforwardly indicated. 
Such a parameter---that makes the action (\ref{S1}) 
non-Lorentz invariant and generates anisotropic spacetimes with a privileged direction---decreases the shadow 
angular radius, as we just read in Fig \ref{Radius}.

\section{Final remarks}
The bumblebee gravity is a Lorentz-violating model in which the VEV of the bumblebee field is nonzero and provides,
for example, a privileged spacetime direction by means of the Lorentz-violating parameter included in the spacetime
metric. Here we built black holes solutions with an effective cosmological 
constant, which comes from a suitable choice for the bumblebee potential. Such an effective cosmological
constant could be either positive or negative. In the first case, we have a Schwarzschild-de Sitter-like
black hole. And for the second case, a Schwarzschild-anti-de Sitter-like black hole. The spacetime 
structure  of both black holes was studied, and the influence of the Lorentz-violating parameter was pointed out,
i.e., that parameter  may increase or decrease the horizons radii depending on the sign of the effective
cosmological constant.

The second part of this article was dedicated to the study of the shadow angular radius in the case of a 
positive and effective cosmological constant, which is the most appropriate case for a cosmological context. 
As we said, the Lorentz-violating parameter influence on 
the shadow angular radius---as far as we know, obtained for the first time in the literature here---decreases 
the shadow angular size.

\section*{Acknowledgments}
We thank Celio Muniz for comments during the article development and an anonymous referee for
valuable points raised in the the review process. 
RVM thanks Fundação Cearense de Apoio ao Desenvolvimento
Científico e Tecnológico (FUNCAP), Coordenação de Aperfeiçoamento de Pessoal de Nível Superior (CAPES), 
and Conselho Nacional de Desenvolvimento Científico e Tecnológico (CNPq, Grant no 307556/2018-2) 
for the financial support.
JCSN also thanks Coordenação de Aperfeiçoamento de Pessoal de Nível Superior (CAPES, Finance Code 001) for
the financial support.


\end{document}